\documentclass[noshowpacs,amsmath,twocolumn,
aps,prb]
{revtex4-1}

\usepackage{setspace}
\usepackage{amsmath}
\usepackage{graphicx}
\usepackage[nearskip,margin = 0pt]{subfig}
\usepackage{subfig}

\usepackage{verbatim}
\usepackage{amsfonts}
\usepackage{amssymb}
\usepackage{epstopdf}
\usepackage{xcolor}
\DeclareGraphicsExtensions{.pdf,.eps,.png,.jpg,.mps}
\usepackage{hyperref}
\hypersetup{
    colorlinks=true,
    linkcolor=blue,
    citecolor=blue,
    filecolor=blue,
    urlcolor=blue,
}

\begin{document}
\title{Quantum diffusion of microcavity solitons}

\author{Chengying Bao,$^{1}$ Myoung-Gyun Suh,$^{1}$ Boqiang Shen,$^{1}$ Kemal \c{S}afak,$^{2}$ Anan Dai,$^{2}$ Heming Wang,$^{1}$ Lue Wu,$^{1}$ Zhiquan Yuan,$^{1}$ Qi-Fan Yang,$^{1}$ Andrey B. Matsko,$^{3}$ Franz X. K$\rm\ddot{a}$rtner,$^{4, 5}$ and Kerry J. Vahala$^{1, *}$\\
$^1$T. J. Watson Laboratory of Applied Physics, California Institute of Technology, Pasadena, California 91125, USA.\\
$^2$Cycle GmbH, Hamburg 22607, Germany.\\
$^3$Jet Propulsion Laboratory, California Institute of Technology, Pasadena, California 91109, USA.\\
$^4$Center for Free-Electron Laser Science, Deutsches Elektronen-Synchrotron, Hamburg 22607, Germany.\\
$^5$Department of Physics and the Hamburg Center for Ultrafast Imaging, University of Hamburg, Hamburg 22761, Germany.\\
$^*$Corresponding author: vahala@caltech.edu
}

\maketitle
\newcommand{\ts}{\textsuperscript}
\newcommand{\tsb}{\textsubscript}


{\bf Coherently-pumped (Kerr) solitons in an ideal optical microcavity are expected to undergo random quantum motion that determines fundamental performance limits in applications of soliton microcombs. 
Here, this diffusive motion and its impact on Kerr soliton timing jitter is studied experimentally. Typically hidden below technical noise contributions, the quantum limit is discerned by measuring counter-propagating solitons. 
Their relative motion features only weak interactions and also presents excellent common mode suppression of technical noise.  This is in strong contrast to co-propagating solitons which are found to have relative timing jitter well below the quantum limit of a single soliton on account of strong mutual motion correlation. Good agreement is found between theory and experiment. The results establish the fundamental limits to timing jitter in soliton microcombs and provide new insights on multi-soliton physics.}



Recently, coherently pumped solitons \cite{Wabnitz_OL1993, Coen_NP2010fiber} have been realized in optical microcavities \cite{Kippenberg_NP2014}. Unlike earlier temporal optical solitons, these new solitons are able to regenerate through Kerr-induced parametric amplification \cite{Vahala_PRL2004Kerr, Maleki_PRL2004}, and strong resonant build-up in the high-Q microcavity enables access to optical nonlinearity at low power levels \cite{Vahala_Nature2003Review}. These desirable features make these devices well suited as chip-scale frequency comb sources or microcombs \cite{Kippenberg_Science2018Review}. The random motion of solitons in these systems (timing jitter) is of central importance in many of their applications. But while the quantum limit of this motion has been studied theoretically \cite{Matsko_OE2013Jitter}, its measurement has not been possible on account of technical noise masking of fundamental fluctuations. It is also unclear if the predicted quantum timing jitter level is achievable in practice.

Here, we experimentally observe the quantum diffusion of microcavity solitons as well their overall timing jitter dynamics. Technical noise suppression has been reported for both counter-propagating (CP) \cite{Vahala_NP2017Counter} and co-propagating (CoP) \cite{Kippenberg_NP2018spatial} soliton pairs in a single microcavity. This suggests that measurement of the relative motion of such a soliton pair could provide a way to observe quantum noise. However, dispersive waves \cite{Menyuk_OL1986,Akhmediev_PRA1995,Kippenberg_Science2016CR,Vahala_Optica2016DW} are known to stabilize the relative positions of CoP solitons \cite{Coen_Optica2017,Matsko_EPJD2017} and to enable the existence of complex structures called soliton crystals \cite{Papp_NP2017Crystal}. These interactions are shown to interfere with the observation of the intrinsic quantum noise associated with a single soliton's motion. On the other hand, CP solitons feature much weaker interactions that rely upon optical backscattering \cite{Vahala_NP2017Counter}, which suggests that observation of weak quantum fluctuations could be possible in these systems. We use this feature of CP solitons to observe the quantum noise limit of soliton motion. Moreover, the different soliton interaction dynamics in CP and CoP systems are also studied. The motions are also numerically examined using the coupled Lugiato-Lefever equations \cite{Coen_OL2013,Vahala_NP2017Counter}.    

\begin{figure*}[t!]
\begin{centering}
\includegraphics[width=\linewidth]{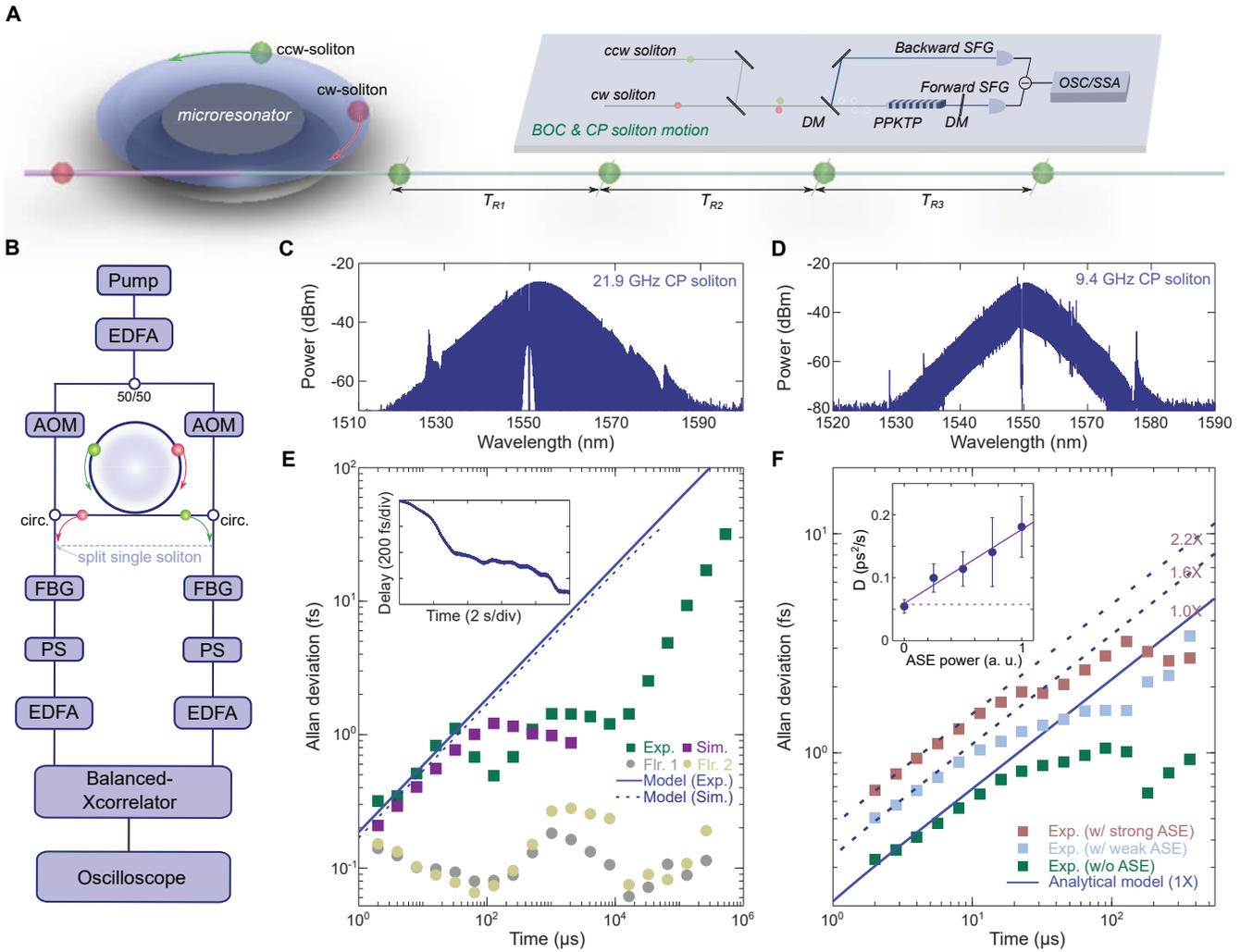}
\captionsetup{singlelinecheck=no, justification = RaggedRight}
\caption{{\bf Counter-propagating solitons and measurmeent of quantum limited motion.} \textbf{A} Solitons undergo random motion in the presence of noise. This can be measured by a balanced optical cross-correlator using counter-propagating (CP) solitons as inputs. DM: dichroic mirror, SFG: sum frequency genenration, OSC: oscilloscope, SSA: signal source analyzer. \textbf{B} Detailed experimental setup for the measurement of CP soliton motion. The dashed line indicates the injection position for the split soliton measurement. AOM: acousto-optical modulators, FBG: fiber Bragg grating, PS: pulse shaper, EDFA: erbium-doped fiber amplifier. \textbf{C, D} Optical spectra for the 21.9 GHz and 9.4 GHz CP solitons. \textbf{E} Allan deviation of the measured (21.9 GHz device) and simulated CP soliton motion. The solid and dashed lines are the predicted Allan deviation from the analytical model using experimentally measured and simulated parameters. The inset shows an example of the measured CP soliton motion on a long time scale. As discussed further in the Supplement, the circle points indicate the measurement floor without (flr. 1) and with an additional 5 m fiber inserted into the one of the fiber paths (flr. 2).  \textbf{F} The Allan deviation of the CP solitons (9.4 GHz device) in the presence and absence of additional ASE. The solid and dashed lines are 1$\times$, 1.6$\times$ and 2.2$\times$ of the analytical model, respectively. The inset summarizes the change of the Allan variance diffusion coefficient (D, see main text) when increasing the ASE power injected into one direction of the pump. The solid line in the inset is a linear fit and the dashed line is the theoretical quantum limited value for D.}
\label{Fig3CP}
\end{centering}
\end{figure*}

\begin{figure*}[t!]
\begin{centering}
\includegraphics[width=0.67\linewidth]{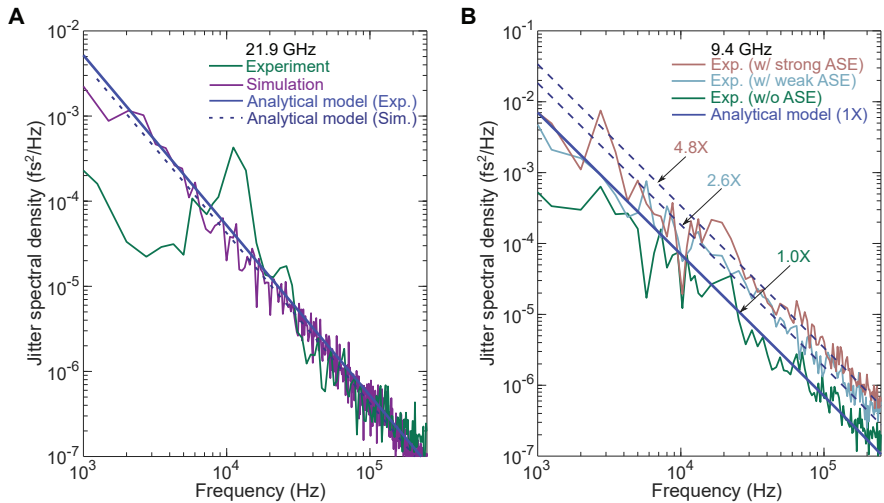}
\captionsetup{singlelinecheck=no, justification = RaggedRight}
\caption{{\bf Measured and theoretical jitter spectral density.} \textbf{A} Measured jitter spectral density of the CP solitons in the 21.9 GHz (green line) and the simulated jitter spectral density (purple line). Both are close to the theoretically predicted jitter spectral density. \textbf{B} The measured jitter spectral density for the 9.4 GHz device for cases without ASE and with added ASE into one propagation direction. The analytical theory is also plotted and the dashed lines are 4.8 and 2.6 times the analytical theory. The theoretical spectral density plotted is 4$\times$ the value of Eqn. \ref{EqOne} to account for two independent solitons and single-side-band experimental and numerical spectral plots.}
\label{Fig2JSpect}
\end{centering}
\end{figure*}

\begin{figure*}[t!]
\begin{centering}
\includegraphics[width=\linewidth]{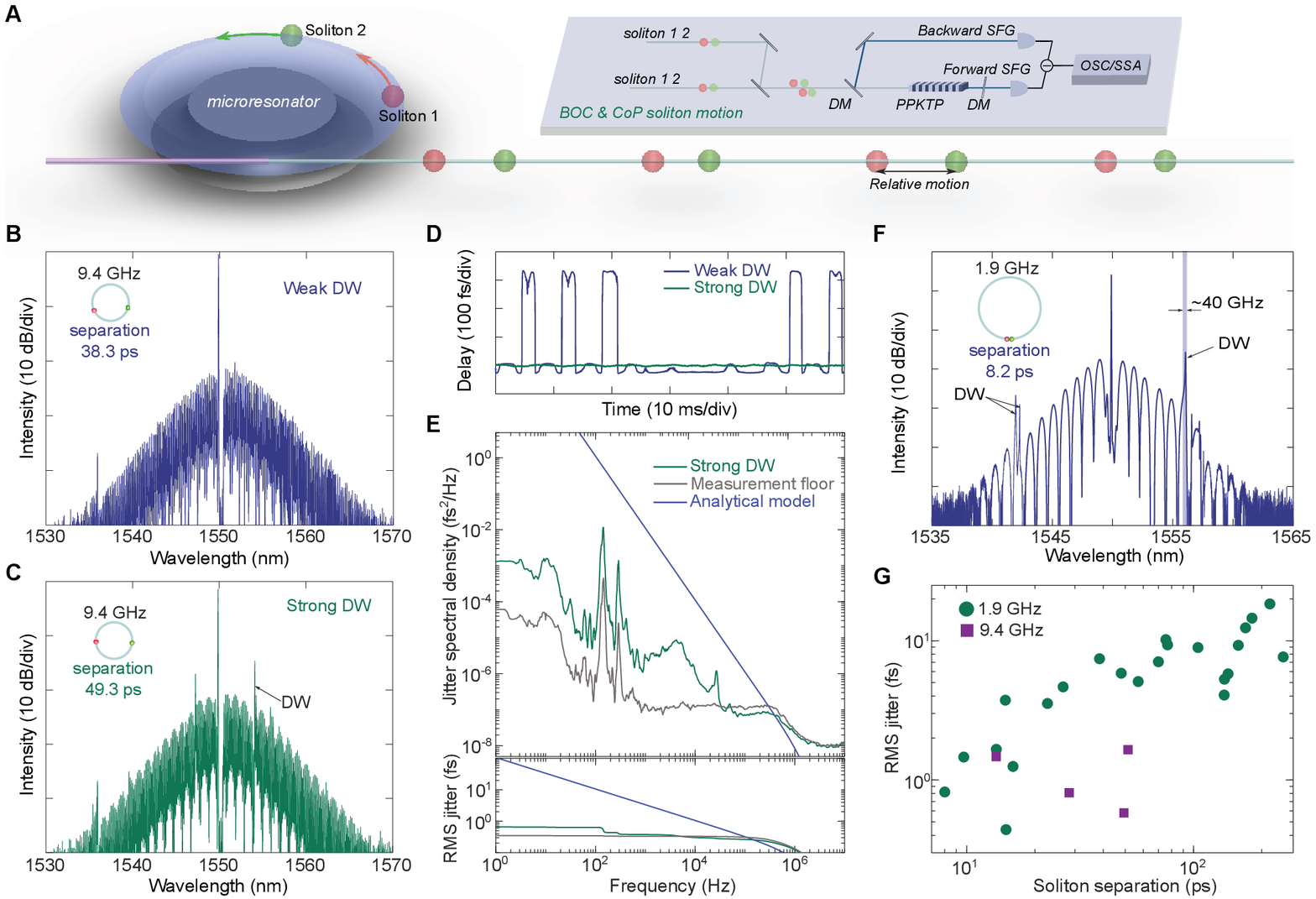}
\captionsetup{singlelinecheck=no, justification = RaggedRight}
\caption{{\bf Relative motion of co-propagating soliton pairs.} \textbf{A} Measurement of relative motion of co-propagating (CoP) solitons. \textbf{B, C} CoP two-soliton optical spectra for a 9.4 GHz device with different dispersive wave strength. The spectral notches near the spectral centers result from an optical filter used to suppress the optical pump. The temporal separation between the solitons is given in the legends. \textbf{D} Relative motion of CoP solitons in the presence of strong and weak dispersive waves. \textbf{E} The jitter spectral density with strong dispersive waves for the relative CoP soliton motion. The blue line and the gray line are the jitter spectral density of the analytical model (zero dispersive wave case) and the measurement noise floor, respectively. Integrated rms relative jitter in the frequency window [1 Hz, 2 MHz] is also shown in this panel. \textbf{F} Two-soliton optical spectrum of a 1.9 GHz resonator with an 8.2 ps separation between 2 solitons. The shaded region is the dispersive waves emission band and contains over 20 modes. \textbf{G} Measured integrated rms relative jitter in the frequency window [1 Hz, 2 MHz] versus soliton separation for the 1.9 GHz and 9.4 GHz cases.}
\label{Fig2Jitter}
\end{centering}
\end{figure*}

Two resonator types were used for CP soliton generation as illustrated in Fig. \ref{Fig3CP}A (3 mm and 7 mm diameters with 21.9 GHz and 9.4 GHz repetition rates, respectively). Coupling to the resonators uses a tapered fiber \cite{Cai2000,Spillane2003} that is in direct contact with the resonator. CP solitons are generated by counter-pumping using a single pump laser as shown in the experimental setup in Fig. \ref{Fig3CP}B. The amplified spontaneous emission (ASE) of the pump was filtered by a 100 GHz optical filter. Counter-pumping frequencies could be adjusted by acousto-optic modulation, however, in the measurement the pump frequencies were equal. The optical spectra of a single CP soliton from each resonator type is presented in Figs. \ref{Fig3CP}C, D. The generated soliton streams were amplified, dispersion compensated by pulse shapers and then conveyed to a balanced optical cross-correlator (BOC) on optical fibers (see Fig. \ref{Fig3CP}A and Supplement for BOC operation). BOC has been used for the characterization of timing jitter in mode-locked lasers with attosecond resolution \cite{Kartner_OL2003,Kartner_LPR2008}.

While CoP solitons are known to feature strong interactions, CP solitons interact more weakly through optical backscattering, which can cause a form of optical trapping \cite{Vahala_Submitted2020}. Nonetheless, as shown here, this weak trapping still permits diffusive transport of the CP solitons over an observable time scale. In the measurement, an oscilloscope was used to record the BOC output signal over the time window during which the CP solitons’ relative delay remains within the BOC operational range (see the inset of Figs. \ref{Fig3CP}E). The Allan deviation, which is normally used for frequency stability evaluation \cite{Barnes_IEEE1971}, is used here instead to analyze the measured relative temporal motion of the CP solitons (see Supplement for the calculation of the Allan deviation). Over a short time scale ($<$30 ${\rm \mu}$s), the calculated Allan deviation ($\sigma$) increases with averaging time ($\tau_A$) and scales as $\sigma^2 \sim$ D $\tau_A$ where D is introduced as a diffusion coefficient. The measured Allan deviation is close to the theoretical prediction (analytical model) based on the quantum-limited diffusive motion of the solitons \cite{Matsko_OE2013Jitter}. The measured Allan deviation is well above the measurement noise floor. This noise floor was characterized by splitting a single soliton train into the two arms of the measurement system as illustrated by the dashed line in Fig. \ref{Fig3CP}B. Two tests of the noise floor (floor 1 and floor 2 in the figure) were performed and are discussed in the Supplement. Numerical simulation of the relative CP soliton quantum motion (see Supplement) also agrees well with the measurement and the analytical model (Fig.  \ref{Fig3CP}E). Such good agreement between the measured timing jitter and the analytical theory was also observed in the 9.4 GHz device (Fig. \ref{Fig3CP}F).

In the data, a roll-over of the Allan deviation is observed with increasing averaging time. This behavior indicates that the quantum-limited soliton diffusion is constrained. Simulations show that this results from weak mutual trapping of the CP solitons that is caused by optical backscattering \cite{Vahala_Submitted2020}. From the data (and simulation), the quantum-noise diffusive behavior is observed when the corresponding temporal fluctuations are much smaller than the trap scale, which can be no smaller than the soliton pulse width (100s of fs).

Ultimately, on a time scale exceeding 10s of ms, another fluctuation behavior is apparent in the Allan deviation data. This drifting-like motion is attributed to forced motion of the CP solitons \cite{Vahala_Submitted2020} driven by random differential variation in the fiber paths that convey the optical pumps. These slow temporal phase changes create slow, random variations in the counter-pump phases and drive relative motion of the CP solitons through the backscatter process. This is analogous to a systematic modulation of relative CP soliton motion that was recently reported using non-degenerate counter-pumps \cite{Vahala_Submitted2020}.

As a further test of the theory and measurements, we injected broadband ASE noise from an independent optical amplifier (i.e., without signal input) into one of the pumping directions and then measured the change of the CP soliton relative motion. This creates non-common-mode noise in the CP soliton motion. Two representative examples of Allan deviation with additional ASE are plotted in Fig. \ref{Fig3CP}F, and these indicate a noisier soliton motion (larger ${\rm D}$) with increasing ASE power. The diffusion coefficient ${\rm D}$ is observed to increase nearly linearly with the input ASE power (the error bars correspond to multiple measurements - more than 5). The y-intercept of the linear fit is also close to the theoretically predicted quantum limit indicated by the horizontal dashed line in the inset of Fig. \ref{Fig3CP}F.

Relative jitter spectral density of CP solitons was also analyzed and compared with the analytical model \cite{Matsko_OE2013Jitter}. This jitter spectral density is obtained by Fourier transform of the temporal motion captured in a 4 ms observation time window and is shown in Fig. \ref{Fig2JSpect}. Consistent with Allan deviation data and analysis, the measured jitter spectral density rolls off as 1/$\omega^2$ and matches the analytical theory at higher offset frequencies. On the other hand, the relative jitter spectral density is suppressed for lower offset frequencies (e.g., $<$10 kHz) in Figs. \ref{Fig2JSpect}A, B, which is consistent with the observed roll-over of the Allan deviation in Fig. \ref{Fig3CP}. Figure \ref{Fig2JSpect}B also shows an increase of the jitter spectral density with increasing input ASE power levels. 
Similarly, the numerically simulated jitter spectral density is also in agreement with the analytical theory (Fig. \ref{Fig2JSpect}A). The close agreement of the measurements with simulations and the analytical theory in Figs. \ref{Fig3CP}, \ref{Fig2JSpect} further confirms that the measured results reflect the quantum limit of soliton motion. These results also suggest that the timing jitter predicted in ref. \cite{Matsko_OE2013Jitter} is achievable with sufficient technical noise suppression.

In contrast to the CP solitons, the CoP solitons are observed to strongly interact via dispersive waves. To study the CoP system, two CoP solitons were generated by controlling the comb power level \cite{Vahala_OL2016}. After dispersion compensation and amplification, the soliton stream generated in a 9.4 GHz microcavity was split and sent into the BOC (Fig. \ref{Fig2Jitter}A). Importantly, dispersive wave strength in this device (different from the one in Fig. \ref{Fig3CP}F) could be controlled by using different segments of the tapered fiber coupler (thicker segments were observed to lead to weaker dispersive waves). Optical spectra for a CoP soliton pair under different dispersive wave strength as well as soliton separation are presented in Figs. \ref{Fig2Jitter}B, C.

Dispersive waves are known to create an effective potential that traps soliton pairs at specific separations \cite{Coen_Optica2017,Matsko_EPJD2017}. In the presence of strong dispersive waves, the relative position of the CoP solitons are constrained in a small range (green line in Fig. \ref{Fig2Jitter}D). In contrast, when the dispersive wave is weaker, the delay between the two CoP solitons can undergo switching between different separations. Similar switching was also reported for CoP soliton pairs in coherently pumped fiber cavities \cite{Coen_Optica2017}. This switching behavior made it difficult to accurately assess fluctuations of the CoP system in the weak dispersive wave case. The relative motion in the strong dispersive waves case is further analyzed as the jitter spectral density (Fig. \ref{Fig2Jitter}E), while the noise floor was also measured by overlapping the CoP solitons with their own replica. The measured jitter spectral density is suppressed by the dispersive wave trapping to a very low level that is several orders lower than the diffusive quantum limit given by the blue line (zero dispersive wave case).

Integrating the jitter spectrum yields the rms jitter (different from the Allan deviation), which is also shown in Fig. \ref{Fig2Jitter}E. An rms jitter of 0.6 fs integrated within the frequency window from 1 Hz to 2 MHz is measured for the strong dispersive wave case. This small number, which is over two orders of magnitude lower than the analytical prediction for two independent (non-interacting) CoP solitons, illustrates the strength of the dispersive wave trapping effect. To our knowledge, this represents the first measurement of jitter spectral density in soliton microcombs showing the interaction of solitons via dispersive waves. 

The length scale of the soliton interaction decreases as the dispersive wave mode number increases \cite{Coen_Optica2017}. To test the impact of this effect on CoP solitons relative motion, the jitter spectral density measurement was repeated many times for two soliton states featuring a wide range of temporal separations within the resonator. The measurements were performed using the 9.4 GHz resonator in Fig. \ref{Fig2Jitter} and a 1.9 GHz resonator. The narrower mode spacing of the 1.9 GHz device more readily accommodates multi-mode dispersive wave emission which is expected to decrease the interaction length scale of the trapping potential \cite{Coen_Optica2017}. A typical optical spectrum for a two soliton state in this resonator is shown in Fig. \ref{Fig2Jitter}F. The shaded region in the spectrum overlaps the dispersive waves and encompasses about 20 resonator modes. The measured integrated rms timing jitter (1 Hz to 2 MHz) versus soliton separation is plotted in Fig. \ref{Fig2Jitter}G (green points) and increases with increasing soliton separation. On the other hand, there is no consistent dependence of the rms jitter on the soliton separation when this measurement was repeated using the 9.4 GHz resonator. This device, in contrast to the 1.9 GHz device,  features very narrow band (nearly single mode) dispersive waves as observed in Fig. \ref{Fig2Jitter}C.

The experimental validation of the quantum theory means that it is possible to consider optimization of microcavity design for minimal quantum jitter. The quantum limited timing jitter spectral density is given by \cite{Matsko_OE2013Jitter},
\begin{equation}
\begin{aligned}
 S_t(\omega) \simeq \frac{\hbar \omega_0 \kappa}{E \omega^2} \Bigg[\frac{\pi^2\tau^2}{12} + \frac{\left|\beta_2\right|^2 v_g^2}{3 \tau^2 \left(\omega^2+\kappa^2\right)} + \frac{\omega \left|\beta_2\right| v_g}{\omega^2+\kappa^2} \Bigg],
\end{aligned}
\label{EqOne}
\end{equation}
where $\omega_0$ is the pump frequency, $E$ is the intracavity pulse energy, $\kappa$ is the total loss rate including both intrinsic loss and coupling loss, $\tau$ is the soliton pulse-width (0.57 of the FWHM pulse-width), $\beta_2$ is the group velocity dispersion, and $v_g=L/T_R$ is the group velocity of the soliton ($L$ is the cavity round-trip length and $T_R$ is the round-trip time). It is straightforward to show that the second term in Eqn. \ref{EqOne} is typically larger than the other terms (see Supplement).
Keeping only this term and considering low offset frequencies ($\omega \ll \kappa$), Eqn. \ref{EqOne} simplifies to the following form,

\begin{equation}
S_{t}(\omega) \approx \frac{\hbar \omega_{0}}{\omega^{2}} \frac{\gamma\left|\beta_{2}\right| v_{g}^{2}}{6 \kappa \tau}.
\label{S1S3}
\end{equation}
where $\gamma=n_2\omega_0/(cA_\text{eff})$ is the nonlinear coefficient ($c$ is the light velocity in the vacuum, $n_2$ is the nonlinear index, and $A_\text{eff}$ is the effective mode area). Therefore, low dispersion mode families (low $|\beta_2|$), weaker nonlinear coefficients, larger modal areas, and larger dissipation modes (large $\kappa$) are preferred to minimize the quantum-limited jitter.  These considerations must be weighed against their collateral impact on pumping power. For example, threshold pumping power will increase approximately quadratically with $\kappa$ and inversely with $\gamma$.

In summary, we have studied microcavity soliton relative motion under both counter-propagating and co-propagating conditions with sub-femtosecond time resolution. The motion of counter-propagating solitons provides a combination of common-mode-noise suppression and independence that enables observation of diffusion-like behavior on a short time scale.  The diffusion coefficient of this motion agrees with the quantum noise level predicted by a theory for coherently pumped solitons \cite{Matsko_OE2013Jitter} as well as by numerical simulation.  Moreover, injection of ASE noise into one of the CP soliton paths so as to introduce non-common-mode noise causes the diffusion coefficient to increase. CoP solitons, on the other hand, feature motion that reflects the influence of soliton pair coupling induced by dispersive wave mediated interactions. Their relative timing jitter was also observed to increase with soliton separation in cases of multi-mode dispersive wave emission while it maintained a low value at all soliton separations for the case of dispersive wave emission on a small number of modes. The analytical theory was also used to consider optimal designs for low timing jitter operation. The mechanisms studied here are expected to be universally observable in soliton microcomb systems generated using other material platforms. The ability to measure soliton timing precisely may also pave the way for study of quantum solitons \cite{Yamamoto_Nature1993} in coherently pumped systems.

\vspace{3 mm}

\noindent \textbf{Acknowledgments}
This work is supported by the Air Force Office of Scientific Research (FA9550-18-1-0353) and the Kavli Nanoscience Institute. CB gratefully acknowledges a postdoctoral fellowship from the Resnick Institute at Caltech. The work of AM was carried out at the JPL, Caltech, under a contract with the National Aeronautics and Space Administration.
\vspace{1 mm}

\noindent\textbf{Author Contributions}
CB, K\c{S}, AM, FXK and KJV conceived the project. CB ran the experiments with assistance from SB, ZY and QFY. ~MGS, HW and LW prepared the samples. ~K\c{S}, AD and FXK built the BOC. The project was supervised by KJV.
\vspace{1 mm}

\noindent \textbf{Author Information} Correspondence and requests for materials should be addressed to KJV (vahala@caltech.edu).

\bibliography{main}
\end{document}


\title{Supplementary Materials for ``Quantum diffusion of microcavity solitons"}

\author{Chengying Bao,$^{1}$ Myoung-Gyun Suh,$^{1}$ Boqiang Shen,$^{1}$ Kemal \c{S}afak,$^{2}$ Anan Dai,$^{2}$ Heming Wang,$^{1}$ Lue Wu,$^{1}$ Zhiquan Yuan,$^{1}$ Qi-Fan Yang,$^{1}$ Andrey B. Matsko,$^{3}$ Franz X. K$\rm\ddot{a}$rtner,$^{4, 5}$ and Kerry J. Vahala$^{1, *}$\\
$^1$T. J. Watson Laboratory of Applied Physics, California Institute of Technology, Pasadena, California 91125, USA.\\
$^2$Cycle GmbH, Hamburg 22607, Germany.\\
$^3$Jet Propulsion Laboratory, California Institute of Technology, Pasadena, California 91109, USA.\\
$^4$Center for Free-Electron Laser Science, Deutsches Elektronen-Synchrotron, Hamburg 22607, Germany.\\
$^5$Department of Physics and the Hamburg Center for Ultrafast Imaging, University of Hamburg, Hamburg 22761, Germany.\\
$^*$Corresponding author: vahala@caltech.edu
}

\maketitle

\renewcommand{\thefigure}{S\arabic{figure}}
\renewcommand{\theequation}{S\arabic{equation}}

\renewcommand*{\citenumfont}[1]{S#1}
\renewcommand*{\bibnumfmt}[1]{[S#1]}
\renewcommand{\thesection}{\arabic{section}}
\renewcommand{\thetable}{S\arabic{table}}

\noindent \textbf{\Large{Materials and Methods}}
\vspace{3 mm}

\noindent \textbf{BOC operation}

The BOC works by balanced detection of forward and backward sum frequency generation signals in a type-\textrm{II} PPKTP crystal. The two optical inputs are adjusted to be orthogonally polarized. When the delay between the two arms is scanned, the BOC outputs an `S'-like voltage signal \cite{Kartner_OL2003,Kartner_LPR2008}. By operating near the zero-crossing region of the `S'-like output, the voltage signal is linear with respect to the pulse delay so that relative delay fluctuations between the two solitons are converted into a voltage change. Two pulse shapers in Fig. 1B are used to compensate optical fiber dispersion so as to enhance the sensitivity the BOC signal. A pulse shaper is also used in the CoP soliton measurement before splitting into two arms. The BOC voltage signal is then measured using either an oscilloscope or a signal source analyzer.

\vspace{3 mm}
\noindent \textbf{Allan deviation noise floor}

The measurement noise floor 1 (flr. 1) in Fig. 1E was determined by generating a single soliton pulse stream in one direction and then splitting it using a bidirectional coupler. The coupler outputs were then connected into the two fiber paths leading to the BOC (see Fig. 1B). These paths include all components except the two optical circulators and this measurement therefore tests the impact of these components on the Allan deviation floor. Moreover, the length of the two paths was estimated to be approximately 38 m and 41 m and the possible impact of propagation through these fiber lengths was therefore also included in this measurement.  Because balancing the propagation paths to the BOC to better than a few meters was difficult, the CP solitons measured by the BOC leave the resonator at different times. Their departure time difference amounts to possibly 100s of resonator round trips. To test any possible impact of this delay on the noise floor, an additional measurement (flr. 2) was performed in which a 5 m long segment of optical fiber was introduced into the longer path. The extra path length had minimal impact on the short term noise floor where the quantum noise is observable. At longer time scales differences were observed that are believed to result from path length changes induced by temperature drift.

\vspace{1 mm}
The 5 m path delay measurement is notable in one other sense. It shows that the impact of this amount of equvalent time delay in pump excitation of the soliton has negligible impact on the short term Allan deviation. While noise floor 2  corresponds to a split soliton measurement, it strongly suggests that such a path difference in excitation of the CW and CCW directions of the CP solitons would have little or no impact on the short term Allan deviation of their relative delay. This is important because the path difference of the two pumping arms is balanced to no better than a few meters.

\vspace{3 mm}
\noindent \textbf{Theoretical jitter spectral density and Allan deviation calculation}

The theoretical timing jitter spectral density presented in Fig. 2A (solid line), Fig. 2B and Fig. 3E is calculated using Eqn. 1 with experimentally measured parameters as following,

\begin{table}[hbtp]
\renewcommand\arraystretch{1.25}
\centering
\begin{tabular}{ccccccc}
\hline
\hline
~~~~~~~ & ~~$T_R$ (ps)~~ & ~~$\beta_2$ (ps$^2$/km)~~ & ~~$L$ (mm)~~ & ~~$\tau$ (fs)~~ & ~~$E$ (pJ)~~ & ~~$\kappa$ (MHz)~~ \\
\hline
Fig. 2A & 46 & $-$21 & 9.4 &  131 & 66 & 2$\pi\times$2.1 \\
Fig. 2B & 107 & $-$33 & 22 & 245 & 100 & 2$\pi\times$0.8 \\
Fig. 3E & 107 & $-$35 & 22 & 188 & 61 & 2$\pi\times$1.5 \\
\hline
\hline
\end{tabular}
\label{Parameters}
\caption{Experimental parameters for the theoretical jitter spectral density plots}
\end{table}

The corresponding Allan deviation for solid lines in Figs. 1E, F is computed using the following equation \cite{Barnes_IEEE1971}
\begin{equation}
\sigma(\tau_A)=\sqrt{2 \int_{0}^{f_{h}} 4 S_{t}(f) \frac{\sin ^{4}(\pi f \tau_A )}{(\pi f \tau_A)^{2}} d f},
\label{Allan}
\end{equation}
where $\tau_A$ is the averaging time, $S_t$ is given in Eqn. 1 and the factor 4 accounts the relative single-side-band jitter spectral density for 2 solitons and $f_h$ is a high frequency chosen as 10 MHz.

\vspace{3 mm}
\noindent \textbf{Simplification of the theoretical model}

Comparing the first term $S_t^{(1)}$ and second term $S_t^{(2)}$ in Eqn. 1 at low offset frequencies $\omega \ll \kappa$ shows that,
\begin{equation}
\frac{S_{t}^{(1)}}{S_{t}^{(2)}}=\left(\frac{\pi \tau^{2} \kappa}{2\left|\beta_{2}\right| v_{g}}\right)^{2}=\left(\frac{\pi \kappa}{4 \delta \omega}\right)^{2} \ll 1,
\label{S1S2}
\end{equation}
where $\delta \omega$ is the detuning of the cavity resonant frequency relative to the pump frequency and is typically much larger than $\kappa$ for stable soliton mode locking. In deriving this result the relationship $\gamma P=\left|\beta_{2}\right| / \tau^{2}=2 \delta \omega / v_{g}$ \cite{Wabnitz_OL1993, Kippenberg_NP2014} has been used where $P$ is the peak soliton power in the cavity and $\gamma=n_2\omega_0/(cA_\text{eff})$ is the nonlinear coefficient ($c$ is the light velocity in the vacuum, $n_2$ is the nonlinear index, and $A_\text{eff}$ is the effective mode area).

Similarly, comparing the second term $S_t^{(2)}$ and third term $S_t^{(3)}$ in Eqn. 1 at low offset frequencies shows that,

\begin{equation}
\frac{S_{t}^{(2)}}{S_{t}^{(3)}}=\frac{\left|\beta_{2}\right| v_{g}}{3 \omega \tau^{2}}=\frac{2 \delta \omega}{3\omega} \gg 1
\label{S2S3}
\end{equation}

Therefore, the theoretical model can be simplified into Eqn. 2 of the main text.

\vspace{3 mm}
\noindent \textbf{Numerical simulations of random motion}

The simulation of quantum noise limited soliton motion uses a similar method as described in \cite{Paschotta_APB2004}. The simulation model is based on two coupled Lugiato-Lefever equations (LLEs), which can be written as \cite{Coen_OL2013,Vahala_NP2017Counter}
\begin{equation}
\begin{aligned}
\frac{\partial A_{1}}{\partial T}=-\left(\frac{\kappa}{2}+i \delta \omega \right) A_{1}-i \frac{\beta_{2} L}{2 T_{R}} \frac{\partial^{2} A_{1}}{\partial t^{2}}+i\frac{\kappa_b}{2} A_{2}+ \sqrt{\frac{\kappa_{e}P_\text{in}}{T_{R}}} + \frac{i \gamma L}{T_{R}} \int_{-\infty}^{+\infty} R\left(t^{\prime}\right)\left|A_{1}(T,t-t^{\prime})\right|^{2} d t^{\prime} +  F_1(t,T),
\end{aligned}
\label{LLE1}
\end{equation}

\begin{equation}
\begin{aligned}
\frac{\partial A_{2}}{\partial T}=-\left(\frac{\kappa}{2}+i \delta \omega \right) A_{2}-i \frac{\beta_{2} L}{2 T_{R}} \frac{\partial^{2} A_{2}}{\partial t^{2}}+i\frac{\kappa_b}{2} A_{1} +\sqrt{\frac{\kappa_{e}P_\text{in}}{T_{R}}}  + \frac{i \gamma L}{T_{R}} \int_{-\infty}^{+\infty} R\left(t^{\prime}\right)\left|A_{2}(T,t-t^{\prime})\right|^{2} d t^{\prime} + F_2(t,T),
\end{aligned}
\label{LLE2}
\end{equation}
where $T, t$ are slow time and fast time, respectively; $A_1$ and $A_2$ are the envelope of the intracavity field in two directions, $\delta \omega$ is the pump-resonator frequency detuning, $\kappa$ is the total loss rate, $\kappa_e$ is the external coupling rate, $\kappa_b$ is the backscattering rate from a point backscatter, and $F_{1(2)}$ is the noise term. The nonlinear response function $R(t)=(1-f_r)\delta(t)+f_r h_R$ includes the instantaneous electronic and delayed Raman contributions. The Raman contribution is calculated in the frequency domain assuming a Lorentzian gain spectrum centered at $-$14 THz and a 3 dB bandwidth of 5 THz, and $f_R$=0.22. The quantum noise is added as white noise across the optical spectrum with a variance of $\langle F_{1(2)}(t,T)F_{1(2)}(t',T')\rangle=100\times \hbar\omega_0\kappa \delta(t-t',T-T')/dt$, where $\omega_0$ is the pump frequency (2$\pi\times$193 THz), and $dt$ is the temporal resolution of the fast time window \cite{Paschotta_APB2004}. The relative soliton position versus slow time $T$ is saved for Allan deviation and jitter spectral density analysis. A factor of 100 is included in the noise force to exaggerate relative soliton motion \cite{Paschotta_APB2004}, and this factor is normalized-out in the subsequent jitter spectral density and Allan deviation calculation.

Other parameters used in simulations are $L$=9.4 mm, $T_R$=40.96 ps, $\kappa=2\pi\times1.9$ MHz, $\kappa_e=2\pi\times0.6$ MHz, $\kappa_b=2\pi\times7.8$ kHz, $\beta_2=-$22 ps$^2$/km, $\gamma$=2.7 W$^{-1}$km$^{-1}$, $P_\text{in}$= 156 mW, $\delta\omega=2\pi\times$29 MHz. The simulated soliton parameters include $\tau$=120 fs and $E$=140 pJ for the calculation of analytical Allan deviation (dashed line in Fig. 1E) and jitter spectral density (dashed line in Fig. 2A).

\bibliography{main_SI}